\documentclass[prb,aps,twocolumn,showpacs]{revtex4-1}
\pdfoutput=1
\usepackage{dcolumn}
\usepackage{bm}

\usepackage[colorlinks,hyperindex, urlcolor=blue, linkcolor=blue,citecolor=black, linkbordercolor={.7 .8 .8}]{hyperref}
\usepackage{graphicx}
\usepackage{amsfonts}
\usepackage{amsmath}
\usepackage{amssymb}
\usepackage{amsbsy}
\usepackage{nicefrac}

\newcommand{\Vae}{$V_{ae}~$}

\begin{document}

\title {Flip-chip gate-tunable acoustoelectric effect in graphene}

\author{J.R. Lane$^1$}
\author{L. Zhang$^1$}
\author{M.A. Khasawneh$^{1}$} 
\altaffiliation[Present address: ]{United States Coast Guard Academy, 31 Mohegan Avenue, New London, CT, 06320, USA}
\author{B.N. Zhou$^2$}
\author{E.A. Henriksen$^{2,3}$} 
\author{J. Pollanen$^1$}
\email[]{pollanen@pa.msu.edu}
\affiliation{$^1$Department of Physics and Astronomy, Michigan State University, East Lansing, MI 48824-2320, USA}
\affiliation{$^2$Department of Physics, Washington University in St. Louis, St. Louis, MO 63130, USA}
\affiliation{$^3$Institute for Materials Science \& Engineering, Washington University in St.~Louis, 1 Brookings Dr., St.~Louis MO 63130, USA}

\date{\today}

\begin{abstract}
We demonstrate a flip-chip device for performing low-temperature acoustoelectric measurements on exfoliated two-dimensional materials. With this device we study gate-tunable acoustoelectric transport in an exfoliated monolayer graphene device, measuring the voltage created as high-frequency surface acoustic waves dynamically drive the graphene charge carriers, the density of which we simultaneously control with a silicon back-gate. We demonstrate ambipolar dependence of the acoustoelectric signal, as expected from the sign of the graphene charge carriers. We observe a marked reduction in the magnitude of the acoustoelectric signal over a well-defined range of density in the vicinity of charge neutrality, which we attribute to a spatially heterogeneous charge-disorder landscape not directly revealed by conventional transport measurements.
\end{abstract}

\maketitle

\section{Introduction}
The electric field co-propagating with a high-frequency piezoelectric surface acoustic wave (SAW) can be coupled to a nearby two-dimensional electron system (2DES). This coupling can be utilized to develop versatile tools for studying and controlling the electronic properties of a 2DES. Surface acoustic wave techniques have been particularly successful for studying 2DESs in GaAs/AlGaAs heterostructures due to the piezoelectricity of the host semiconductor. In these systems SAW techniques have been used to investigate the quantum Hall regime\cite{wix86, wix89, ess92, kuk11, pol16, fri17}, Wigner crystallization at low carrier density\cite{paa92, dri15, sus15}, the existence of a composite fermion metal near Landau level filling factor $\nu = 1/2$\cite{wil93, wil932}, and the 2D metal-to-insulator transition at zero magnetic field\cite{tra06}. The extension of these techniques to exfoliated two-dimensional materials is in its nascency and is opening the toolkit of high-frequency acoustoelectrics to a host of new low-dimensional electron systems.

In particular, in recent years theoretical\cite{tha10, zha11, sch13, ins15, Dom17} and experimental\cite{mis12, mis12t, ban13, san13, May14, Ban15, Poo15, ban16, Zhe16, oku16, poo17, Tan17, Lia2017} efforts have been made to extend SAW techniques to study the wave-vector and frequency dependent electronic properties of graphene and to develop acoustoelectric devices for a variety of quantum applications. However, technical challenges arise when applying these methods to graphene, or other layered two-dimensional materials, that are absent from similar GaAs experiments. The most pressing is the need to incorporate a compatible gate electrode to control the type of charge carriers and their density. The ability to tune charge carriers \textit{in situ} allows for not only SAW studies of fragile quantum electronic states similar to those conducted in GaAs quantum wells, but also the development of hybrid devices for high-frequency charge pumping\cite{shi962, fle03} or for controlling the quantum dynamics of single electronic states\cite{kat09}. 

Oxidized silicon, the most common substrate for graphene devices, is not piezoelectric, and the most strongly piezoelectric materials are highly insulating, making gating graphene devices on these substrates difficult. To overcome this obstacle several studies\cite{oku16, ban16, Tan17} have employed an ion gel gate to demonstrate gate-tunable acoustoelectric transport. Ion gel gating, however, cannot be tuned \emph{in situ} at the low temperatures needed for a variety of quantum transport experiments. Studies interfacing two-dimensional MoS$_{2}$ with SAWs have used an electrode on the back of a peizoelectric substrate to gate the device\cite{pre15}, however the relatively thick gate dielectric in this configuration renders it difficult to tune the carrier density to large values easily. Recently, a surface micromachining aluminum nitride technique was used to integrate a SAW delay line with a graphene field-effect transistor on silicon\cite{Lia2017}, however ambipolar acoustoelectric transport was not observed. While these previous reports all provide methods by which gate tunability can be incorporated it is nonetheless useful to explore other possible routes by which it can be achieved.

In this paper, we report on an alternative method by which both gate-tunability and acoustoelectric measurements can be incorporated simultaneously to study two-dimensional materials at low temperatures. In particular, we demonstrate a gate-controlled acoustoelectric effect in an exfoliated graphene device fabricated directly on an oxidized silicon substrate. To achieve acoustoelectric coupling, we use a flip-chip device geometry where SAWs propagate on a separate piezoelectric substrate which is flipped upside-down and mechanically clamped onto a silicon substrate holding the graphene. We characterize the performance of this device by measuring the gate-dependent acoustoelectric voltage generated as the nearby SAW evanescently couples to graphene charge carriers. We observe a clear dependence of this acoustoelectric voltage as we tune the sign of the charge carriers in the graphene. At low carrier density, we observe a reduction in the magnitude of the acoustoelectric voltage where the signal goes to zero on average over a range of gate voltage values near to graphene charge neutrality. We attribute this behavior to the existence of spatially inhomogeneous charge disorder.

\section{SAW-2DES Interaction and Acoustoelectrics}
As a surface acoustic wave propagates on a piezoelectric substrate, it carries with it an electric field that can interact with a nearby 2DES, coupling the systems in a manner dependent on the properties of the 2DES as well as the frequency, wavelength and intensity of the piezoelectric wave. This interaction arises when the charge carriers react to dynamically screen the electric field of the SAW. During this process, energy and momentum are transferred from the SAW to the low-dimensional charge carriers, which in turn attenuates and shifts the velocity of the piezoelectric wave. In the presence of the two-dimensional carriers the SAW attenuation per unit length $\Gamma$ and its velocity shift $\Delta v/v$ are given by 
\begin{equation}
\Gamma = k\frac{K^2}{2}\frac{(\sigma(k,\omega)/\sigma_m)}{1+(\sigma(k,\omega)/\sigma_m)^2}
\end{equation}
\begin{equation}
\frac{\Delta v}{v} =  \frac{K^2}{2}\frac{1}{1+(\sigma(k,\omega)/\sigma_m)^2}.
\end{equation}
Here, $\sigma(k,\omega)$ is the sheet conductivity of the 2DES at wavenumber $k$ and frequency $\omega$ of the SAW, $K^2$ is a material and device dependent piezoelectric coupling constant, and $\sigma_m \propto \epsilon_r \epsilon_o v$ is a characteristic conductivity that depends on material parameters\cite{wix89}.  The dependence of the attenuation and velocity on $\sigma(k,\omega)$ make SAWs a contactless probe of the sheet conductivity throughout the interior of the 2DES, especially in the regime where $\sigma \simeq \sigma_m$ and both $\Gamma$ and $\Delta v/v$ are maximally sensitive. The piezoelectric substrate used in our flip-chip device is an oxidized form of lithium niobate (LiNbO$_3$ ``black'') for which $\sigma_m$ is of the order of $10^{-6}~\Omega^{-1}$.

The momentum transfer between the SAW and the charge carriers can also drive a macroscopic current in the 2DES\cite{efr90}. This acoustoelectric interaction is inherently nonlinear\cite{efr90,ess93} and in the absence of a magnetic field\cite{shi95, rot98} the acoustoelectric current density in the direction of the SAW propagation is given by
\begin{equation}
j = \sigma E - \frac{\mu I \Gamma}{v}
\end{equation}
where $E$ is the acoustoelectrically generated electric field in the 2DES, $\mu$ is the charge carrier mobility at zero magnetic field, $I$ is the intensity of the surface acoustic wave, $v$ is the SAW velocity, and $\Gamma$ is the SAW attenuation per unit length from Eqns. (1) and (2). Depending on the experimental setup this type of acoustoelectric effect can be detected by measuring either a net flow of SAW driven dc current in a closed circuit or the accumulation of an acoustoelectric voltage, $V_{ae}$, in an open circuit configuration. These techniques are complementary to direct measurements of the SAW velocity shift or attenuation, particularly when the devices under test are micron-scale or smaller. Acoustoelectric measurements of this type have been used in a wide variety of low-dimensional and nano-scale electronic devices ranging from nanoribbons\cite{poo17} to coupled quantum dots\cite{Ebb05}.   
  
\section{Experiment}
\begin{figure}
\begin{center}
\includegraphics[width = 0.85\linewidth]{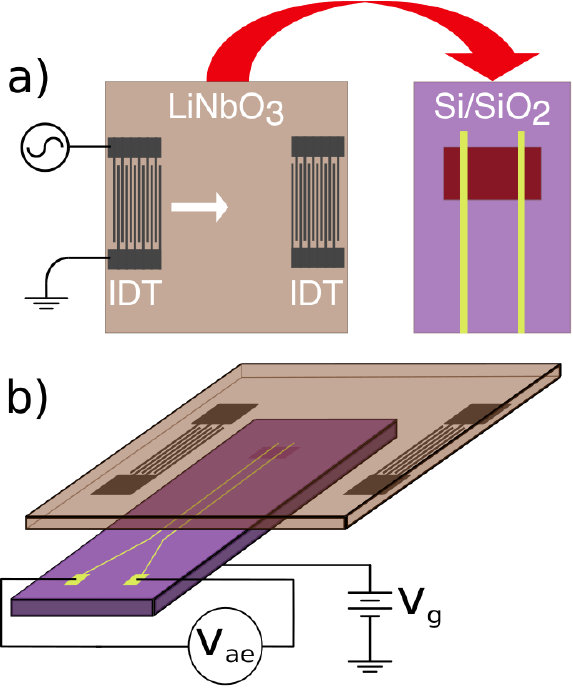}
\end{center}
\caption{(Color online) Schematic of the experimental setup. (a) The flip-chip configuration for the acoustoelectric measurements consists of two devices: a SAW delay line fabricated onto a lithium niobate ``black'' substrate, and an exfoliated graphene device on an oxidized silicon chip. A high-frequency signal applied to one of the interdigitated transducers forming the delay line launches a SAW in the direction indicated by the white arrow. (b) The SAW delay line is flipped onto and mechanically clamped to the graphene device, as described in the text. In this configuration the electric field propagating in tandem with the SAW evanescently couples to the graphene device, while the silicon back gate may be used to control the graphene charge carrier density \textit{in situ}. The resulting acoustoelectric voltage is measured between two electrical contacts that extended across the entire width of the graphene sample.}
\label{fig1}
\end{figure}

The monolayer graphene sample used in this experiment was mechanically exfoliated onto a degenerately doped silicon wafer with a 300 nm SiO$_{2}$ surface layer. Electrical leads for conventional and acoustoelectric transport were fabricated such that they crossed the entire width of the graphene sample perpendicular to the SAW propagation direction as shown in Fig.~\ref{fig1}(a). These leads were defined using standard lithographic techniques, and had a width of $3~\mu$m. Metallization of these electrical contacts was achieved by thermal evaporation of 4 nm of chromium followed by 70 nm of gold. A voltage applied to the entire silicon substrate was used to gate-control the graphene charge carrier density \textit{in situ} as shown in Fig.~\ref{fig1}(b).

Silicon is not piezoelectric and hence SAW measurements cannot be done using a silicon substrate alone. Therefore, we employed a flip-chip configuration, where a pair of interdigitated transducers (IDTs) for launching SAWs were fabricated onto a separate piezoelectric substrate that was then mechanically pressed onto the silicon substrate hosting the graphene device, as shown in Fig.~\ref{fig1}(b). In our device, aluminum IDTs were fabricated in the form of a SAW delay line onto a single-crystalline chip of $128^{\circ}$ Y-rotated lithium niobate ``black'' via conventional photolithography\footnote{In this geometry the SAW propagation direction is in the crystallographic \emph{x}-direction.}. The IDT pair had a center-to-center spacing of 7 mm with each transducer composed of 40 pairs of $3~\mu$m wide interdigitated electrode fingers. 

A SAW is launched along the surface of the lithium niobate by applying an alternating voltage between the two electrodes of an IDT at the fundamental frequency of the transducers, $f = v/\lambda \simeq 332$~MHz. This frequency corresponds to a SAW wavelength $\lambda \simeq 12~\mu$m. The electric field accompanying the SAW evanescently extends above the surface of the lithium niobate, meaning the SAW may couple to a 2DES at a height $< \lambda$ above the lithium niobate \cite{sch88,Kin71}. To effectively couple the SAW to the charge carriers in the graphene, care was taken to ensure the absence of large debris on either substrate that could become lodged between them after being flipped into contact. The two chips were mounted in a custom spring-loaded sample holder to maintain their contact at low temperature and to immobilize their relative lateral motion. Previous studies\cite{sch88, wix89, wix88, key15} of flip-chip assemblies similar to ours have estimated air gaps substantially smaller than the SAW wavelength in our experiment. Therefore, while we are not able to directly measure the air gap in our device, we expect it to be sufficiently small to enable acoustoelectric coupling.

With this experimental setup, low temperature acoustoelectric measurements were made between pairs of leads on the graphene by applying an amplitude modulated signal to one SAW transducer and detecting the correspondingly generated acoustoelectric voltage \Vae using standard ac lock-in techniques. These measurements correspond to the open configuration in which the total acoustoelectric current density $j = 0$ (see Eqn.~(3)) and \Vae arises from the charge displacement produced by the SAW in the direction of its propagation. We note that the electrical leads used for detected \Vae also allow us to perform conventional lock-in based low-frequency (13~Hz) transport measurements to characterize the graphene sample and provide a point of reference when interpreting our acoustoelectric data. Finally, all data were taken in zero magnetic field and $T \simeq 3.2$~K.

\section{Results and Discussion}
\subsection{SAW delay line characterization}
\begin{figure}
\begin{center}
\includegraphics[width=1 \columnwidth]{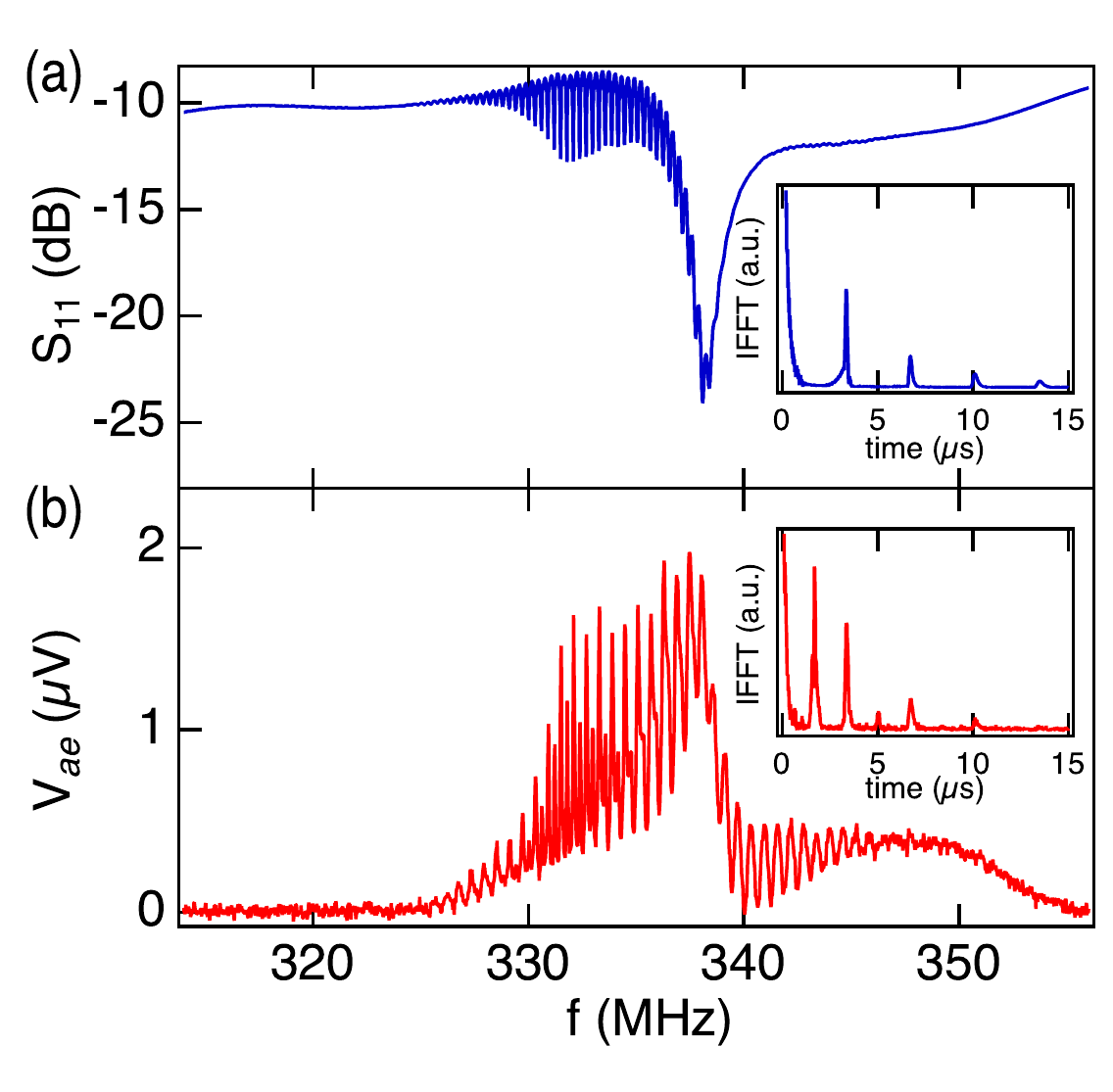}
\caption{(Color online) (a) Frequency dependence of the reflection coefficient ($S_{11}$) of the SAW delay line. Inset: inverse fast Fourier transform (IFFT) of $S_{11}$ data. (b) Measured acoustoelectric voltage in the graphene sample as a function of frequency. This measurement was conducted at 3.2 K with the sample back gate grounded. To excite the SAW, 0 dBm of microwave power was applied at the top of the cryostat. Inset: inverse Fourier transform of the acoustoelectric signal.}
\label{fig2}
\end{center}
\end{figure}

The frequency response of the SAW transducers in our flip-chip device was characterized using an Agilent N5230A vector network analyzer. Fig.~\ref{fig2}(a) shows the measured reflection coefficient $S_{11}$ of the SAW device as a function of frequency. The resonance in the reflected power observed at $\simeq$~337~MHz is associated with the generation of SAWs in the LiNbO$_3$. The slight shift in frequency relative to the expected SAW resonance at 332~MHz is attributable to an increase in the elastic moduli of LiNbO$_3$ and mechanical strain in the flip-chip device upon cooling to cryogenic temperatures\footnote{We note that mass loading alone from the SiO$_{2}$/Si substrate would tend to reduce the resonant response of the SAW delay line. However, since the resonance frequency is shifted slightly higher by $\sim5$ MHz, we conclude that stiffening of the lithium niobate crystal due to strain and cooling to low temperatures dominate mass loading by the SiO$_{2}$/Si substrate.}. The primary SAW resonance shown in Fig.~\ref{fig2}(a) also exhibits superimposed oscillations as a function of frequency. Examining the inverse Fourier transform of $S_{11}$, shown in the inset of Fig.~\ref{fig2}(a), reveals a series of decaying peaks spaced by $3.4~\pm~0.1~\mu$s. This time scale corresponds to a SAW propagation distance of $13.3~\pm~0.2~$mm, which is approximately twice the spacing of the two transducers in our device. Because the transducers also act to reflect surface acoustic waves, a launched SAW will propagate across the chip, be reflected by the opposing transducer, and propagate back to the launching transducer, where its co-propagating electric field will interfere with the reflected electric signal from the launching transducer as measured by the network analyzer. We therefore attribute these superimposed oscillations in $S_{11}$ to the interference between the signal reflected from the launching transducer and the electric field associated with SAWs that have propagated to and from the opposing transducer. 

\subsection{Graphene acoustoelectrics}
Fig.~\ref{fig2}(b) shows the acoustoelectric voltage measured between two leads $29~\mu$m apart\footnote{Consistent behavior was observed using the other leads on the device.} on the graphene device as a function of frequency. The resonance in \Vae coincident with the generation of SAWs centered at $\simeq$~337~MHz is associated with the acoustoelectric transport of charge in the graphene sample. For this data the silicon back gate was held at ground potential and, as we will describe below, the  sample displays p-type conduction typical of graphene devices on SiO$_2$ exposed to air and polymer resist. The sign of the induced voltage is consistent with conduction by holes. 

The frequency response of \Vae shows oscillations superimposed onto the main peak similar to those observed in $S_{11}$. These oscillations are attributable to the modulation in the SAW amplitude caused by interference between the primary SAW and reflected waves.\cite{Ast06} To quantitatively understand the nature of these oscillations, we examine the inverse Fourier transform of $V_{ae}$, which is shown in the inset of Fig.~\ref{fig2}(b). Similar to the behavior of $S_{11}$ we observe a series of decaying peaks spaced by $\simeq 3.4~\mu$s.We attribute these peaks to the interference of the primary SAW with forward propagating SAWs that have been reflected by the far transducer and then again by the original transducer. Additionally, we observe a number of peaks at odd multiples of $1.7~\pm~0.1~\mu$s, which correspond to the propagation distance between the two transducers. We attribute these peaks to modulation in the SAW amplitude caused by counter propagating waves reflected by the opposing transducer. 
\begin{figure}
\begin{center}
\includegraphics[width=1 \columnwidth]{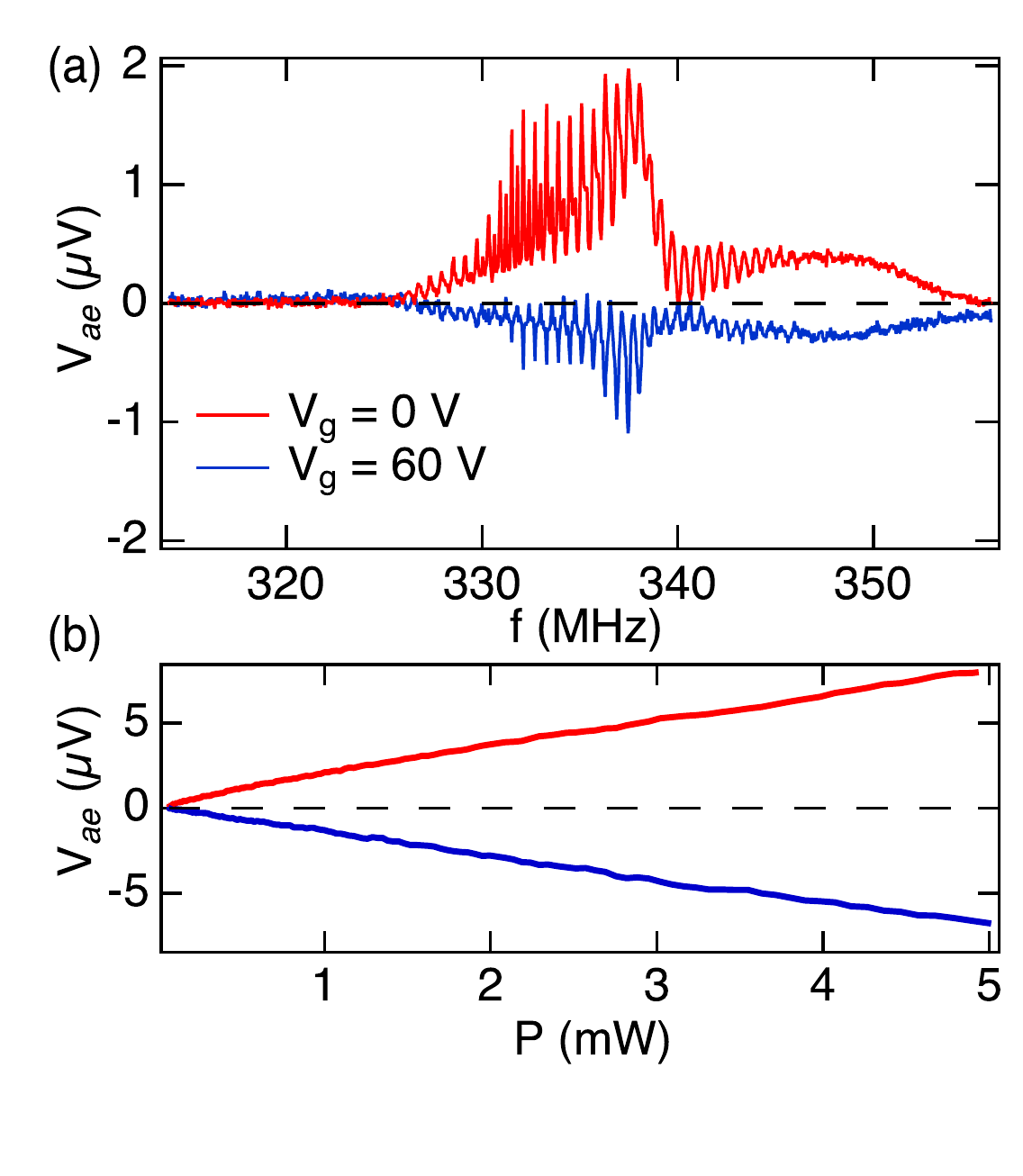}
\caption{(Color online) (a) Acoustoelectric voltage versus frequency for two gate voltages: V$_g$~=~0~V (hole-doped graphene) and V$_g$~=~60~V (electron doped graphene) at 3.2 K. For these measurements 0 dBm of microwave power was used to excite SAWs. (b) Acoustoelectric voltage as a function of power applied to the SAW circuit, for the same back gate voltages as in panel (a). This data was taken at a fixed frequency $f~=~337.45$ MHz corresponding to the maximum in the response of the SAW transducer.}
\label{fig3}
\end{center}
\end{figure}

\begin{figure}
\begin{center}
\includegraphics[width=1 \columnwidth]{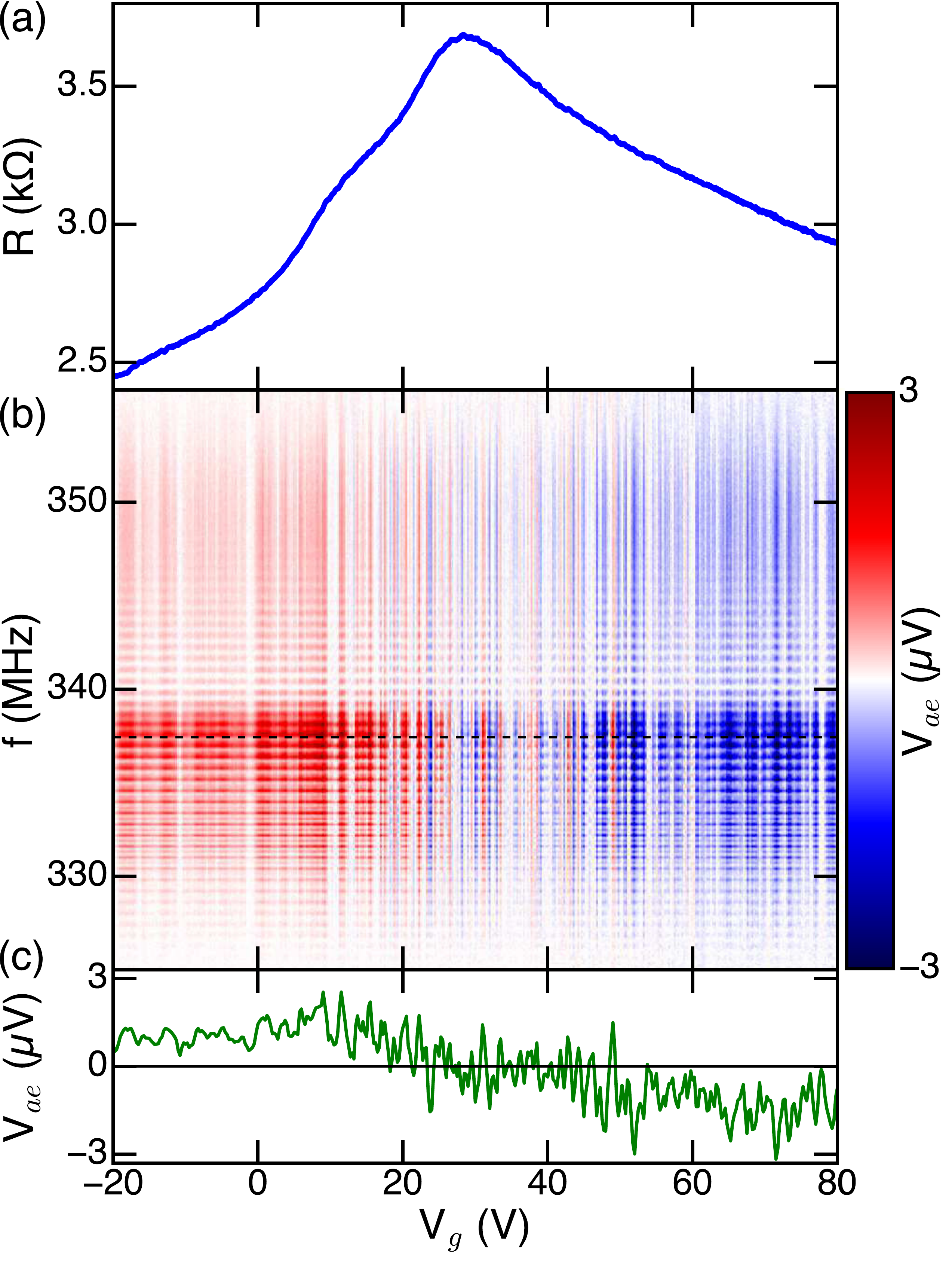}
\caption{(Color online) (a) Low-frequency two-point resistance of the graphene sample as a function of back gate voltage. (b) Acoustoelectric voltage signal (taken with 0 dBm of SAW power), plotted as a function of both back gate voltage and SAW frequency. At gate voltages where the graphene is heavily electron (hole) doped, \Vae is consistently negative (positive). (c) Constant frequency linecut of \Vae at $f~=~337.45$~MHz, indicated by the dashed line in panel (b).}
\label{fig4}
\end{center}
\end{figure}

To verify that the measured voltage is associated with the charge carriers in the graphene, we applied a voltage $V_g$ to the silicon back gate to tune the graphene charge carrier type from holes to electrons. Fig.~\ref{fig3}(a) shows the acoustoelectric voltage as a function of frequency for two back gate voltages on either side of the Dirac peak (see Fig.~\ref{fig4}(a)). With the back gate grounded ($V_{g}=0$~V, red trace Fig.~\ref{fig3}(a)), the graphene exhibits conduction due to holes and the corresponding acoustoelectric signal is positive. By increasing the $V_g$ to +60~V, electrons become the predominant charge carrier in the graphene and, as expected, the acoustoelectric voltage correspondingly reverses sign (blue trace in Fig.~\ref{fig3}(a)). In Fig.~\ref{fig4} we show the full gate- and frequency-dependent map of \Vae for the device along with the corresponding gate-dependent two-point resistance of the graphene. In the bottom panel of Fig.~\ref{fig4} we show a constant frequency linecut of \Vae at $f=337.45$~MHz. As expected, the sign of acoustoelectric signal changes upon tuning through the Dirac peak at $V_{g}~\simeq~28$~V, validating that \Vae is a result of acoustoelectrically induced charge transport in the graphene.
Moreover, in the open configuration of our measurement (where $j = 0$) Eqn. (3) predicts that \Vae should be proportional to the SAW intensity $I$. Therefore, to further verify the acoustoelectric origin of the signal, we measured \Vae as a function of SAW power at a fixed frequency of $f=337.45$~MHz, which corresponds to the peak in SAW resonant response. The results are plotted in Fig. 3(b). For both hole and electron doping we find that \Vae is linear in the applied SAW power, consistent with Eqn. (3) and with previous acoustoelectric measurements in graphene\cite{mis12, ban13, san13, ban16}.

\begin{figure}
\begin{center}
\includegraphics[width=1 \columnwidth]{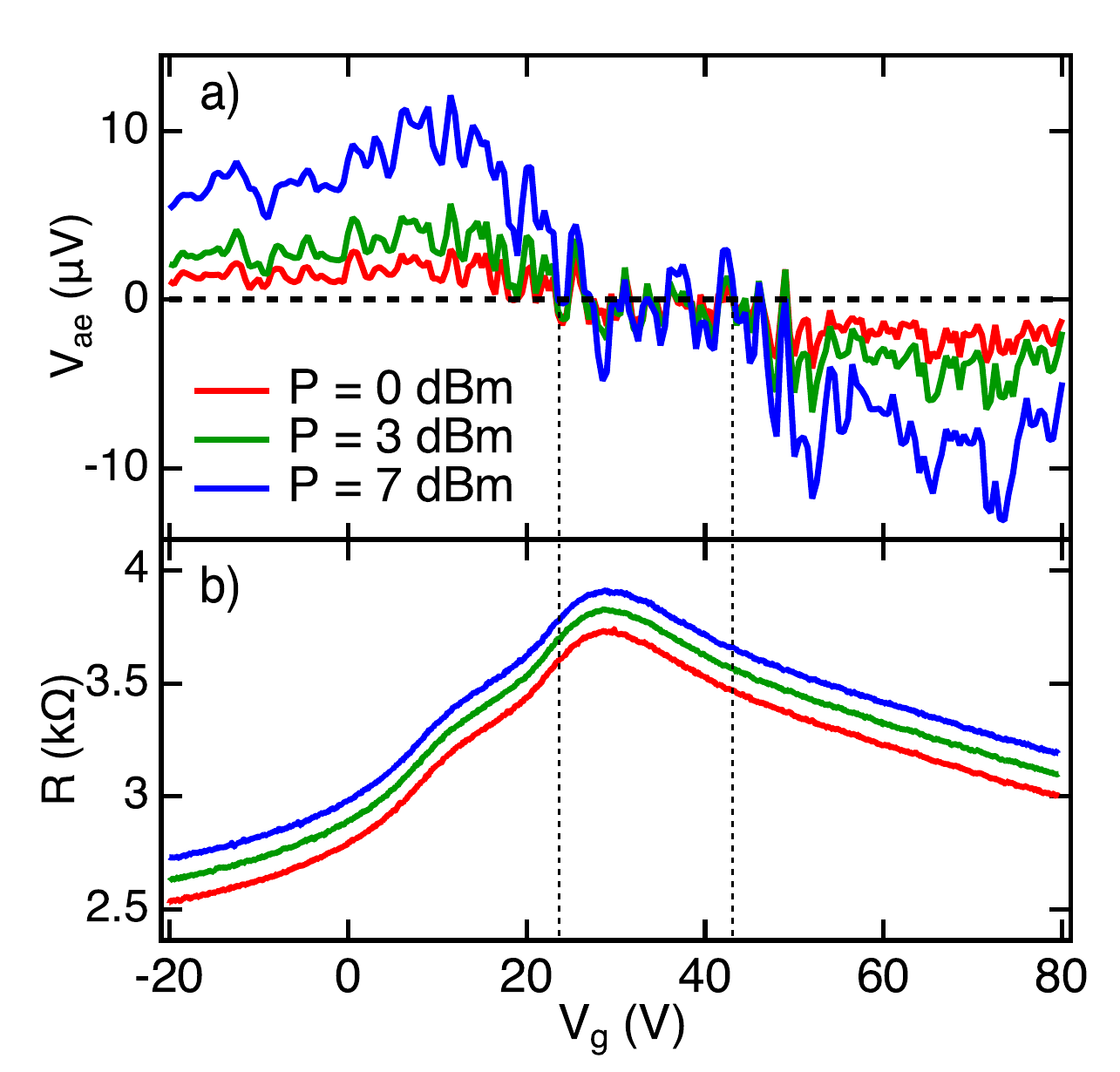}
\caption{(Color online) (a) Acoustoelectric signal versus back gate voltage at a SAW frequency $f~=~337.45$~MHz for increasing levels of SAW power. A well-defined region where \Vae = 0 is observed around charge neutrality and is associated with heterogenous charge disorder as discussed in the text. (b) Corresponding low-frequency two-point resistance of the graphene sample as a function of back gate voltage for the same values of SAW power as shown in (a). The two-point resistance traces in have been offset by 100 $\Omega$ relatively to each other for clarity.}
\label{fig5}
\end{center}
\end{figure}
In the vicinity of the Dirac peak we observe a marked reduction in the magnitude of $V_{ae} \rightarrow 0$ over a well-defined region in gate voltage. This region is emphasized in Fig.~\ref{fig5}(a), where we performed measurements with increasing levels of SAW excitation power to enhance the magnitude of the acoustoelectric signal. For comparison in Fig.~\ref{fig5}(b) we show low frequency transport over the same range of gate voltage while applying a SAW driving signal of the same power. Over a range of $\sim 20$~V, as indicated by the dashed vertical lines, we find that the average value of $V_{ae} \simeq 0$. This range of gate voltage corresponds to a charge carrier density range of approximately $1.4\times10^{12}$~cm$^{-2}$. This region of suppressed \Vae is likely associated with the formation of heterogeneously doped electron-hole ``puddles'', which are known to exist close to charge neutrality due to substrate contamination and impurities \cite{mar08, Li13} but only manifest indirectly in transport measurements as a broadening of the Dirac peak. In this charge disordered regime one would expect that competing acoustoelectric signals from roughly equal distributions of electrons and holes to cancel each other and result in an approximately net zero value of \Vae as observed in our measurements. From the low frequency transport we determine that the field effect mobility of the graphene is $\mu \simeq$~320~cm$^2$/Vs on the hole-doped side of the Dirac peak and $\simeq$~110~cm$^2$/Vs on the electron-doped side. While these values are relatively low for an exfoliated device they are, however, two orders of magnitude higher than those reported in ion-gated devices used in recent acoustoelectric measurements\cite{ban16, oku16}. Moreover, the fact that the mobility of holes in our sample is roughly three times larger than that of electrons naturally explains the asymmetric position of the region over which $V_{ae} \simeq 0$ relative to the Dirac peak. Finally we note that a reduction in SAW driven \emph{current} was observed near charge neutrality in recent measurements of CVD graphene at room temperature\cite{ban16, oku16, Tan17} using ion gel gating. In these measurements the authors suggest similarly that electron-hole puddling is responsible. Our measurements using a different experimental setup and an exfoliated device at low temperatures are consistent with this interpretation and further highlight that acoustoelectric methods provide complementary methods for investigating 2DES and reveal features of the underlying phenomena not directly manifest in conventional transport.

\section{Conclusion}
In summary, we have demonstrated a flip-chip device for piezo-acoustoelectric measurements of exfoliated two-dimensional materials compatible with gate tunability and low temperature experiments. We have characterized both the high frequency acoustic and acoustoelectric performance of this device by comparing to the well-known electrical behavior of exfoliated graphene. Specifically, we have measured the carrier density dependent acoustoelectric effect in a silicon back-gated graphene device in which the SAW driven acoustoelectric voltage is sensitive to the type of charge carrier present in the sample. Near charge neutrality we have observed a substantial range of density over which the acoustoelectric signal is strongly supressed, a phenomenon attributable to the existence of patches of charge disorder in the graphene that are not directly detectable with conventional low-frequency transport but have been demonstrated with local-probe techniques. The hybrid flip-chip device developed for these measurements allows for high-frequency surface acoustic wave techniques to be applied to the already well-established methods for exfoliation and fabrication of layered two-dimensional devices on oxidized silicon. We expect that similar methods will be employed to study and control the carrier dynamics in extremely high mobility graphene as well as other low-dimensional exfoliated electronic materials.

\begin{acknowledgements}
We thank H. Byeon, S. Hemmerle, K. Nasyedkin, A. Turnbull, and J. Milem for helpful discussions. The Michigan State portion of this work was supported by the Cowen Endowment. B.Z. and E.H. acknowledge support from the Institute of Materials Science and Engineering at Washington University in St. Louis, and the NSF under DMR-1810305.
\end{acknowledgements}

\end{document}